# All optical detection of picosecond spin-wave dynamics in two-dimensional annular antidot lattice

Nikita Porwal[1], Sucheta Mondal[2], Samiran Choudhury[2], Anulekha De[2], Jaivardhan Sinha[2], Anjan Barman[2*] and Prasanta Kumar Datta[1*]

[1]*Department of Physics, Indian Institute of Technology Kharagpur, W.B. 721302, India*
[2]*Department of Condensed Matter Physics and Material Sciences, S. N. Bose National Centre for Basic Sciences, Block JD, Sector III,*
*Salt Lake, Kolkata 700 106, India*
*\*Email of the corresponding authors:* pkdatta@phy.iitkgp.ernet.in, abarman@bose.res.in

**Abstract**

Novel magnetic structures with precisely controlled dimensions and shapes at the nanoscale have potential applications in spin logic, spintronics and other spin-based communication devices. We report the fabrication of two-dimensional bi-structure magnonic crystal in the form of embedded nanodots in a periodic $Ni_{80}Fe_{20}$ antidot lattice structure (annular antidot) by focused ion-beam lithography. The spin-wave spectra of the annular antidot sample, studied for the first time by a time-resolved magneto-optic Kerr effect microscopy show a remarkable variation with bias field, which is important for the above device applications. The optically induced spin-wave spectra show multiple modes in the frequency range 14.7 GHz to 3.5 GHz due to collective interactions between the dots and antidots as well as the annular elements within the whole array. Numerical simulations qualitatively reproduce the experimental results, and simulated mode profiles reveal the spatial distribution of the spin-wave modes and internal magnetic fields responsible for these observations. It is observed that the internal field strength increases by about 200 Oe inside each dot embedded within the hole of annular antidot lattice as compared to pure antidot lattice and pure dot lattice. The stray field for the annular antidot lattice is found to be significant (0.8kOe) as opposed to the negligible values of the same for the pure dot lattice and pure antidot lattice. Our findings open up new possibilities for development of novel artificial crystals.

## 1. Introduction

Structuring of magnetic thin films with precisely controlled dimensions and shapes, opens up a unique opportunity to design and prepare magnetic devices such as magnetic random access memory [1], Hall sensors [2], biosensing devices [3], magnetic logic devices [4,5] and magnetic resonance imaging [6]. Thus, it is interesting to observe how the structuring influences the spin-wave (SW) dynamics to tune the features of the magnetic devices. Periodically patterned ferromagnetic nanostructures have already shown interesting SW transmission properties [7,8]. The vigorous interest in SWs, defined as coherent precession modes of magnetic moments is due to their great potential in applications [9-11]. The periodicity should be of the range of wavelength of the SW in order to select suitable frequency band. Such materials behave similar to photonic [12] and phononic crystals [13] and are, therefore, called magnonic crystal (MC) – their magnetic counterpart [9-11, 14-16]. An extensive work on the SW dynamics has been reported on dot [17-22] and antidot [23-27] lattices with varying size, shape, lattice constants and symmetry, and also by changing the constituent magnetic materials. Spin-wave propagation properties through ferromagnetic dots [28] and antidots [29] showed their potential in all-magnetic logic and communication devices. Magnonic band structures can be more efficiently controlled in bi-component magnonic crystals (BMCs), where two different materials provide the much-needed material contrast to the propagating SWs as shown theoretically by Vassuer et al [30]. Later a number of studies on one-dimensional and two-dimensional BMCs have been reported where nanostripes or nanodots of two different ferromagnetic materials are placed next to each other [31-33] or a ferromagnetic material is embedded into another ferromagnetic matrix [34-36]. In addition to the better tunability of magnonic band structure, a larger propagation velocity of SWs has been demonstrated due to the presence of direct exchange coupling between the two materials in a filled antidot lattice [36]. Tunable exchange bias-like effects in magnetostatically-coupled 2D hybrid composites are studied earlier by A. Hierro-Rodriguez et al. [37]. Also, Yu et al. demonstrated that a grating coupler of periodically nano-structured magnets provokes multidirectional emission of short-wavelength SWs with enhanced amplitude compared with a bare microwave antenna [38]. The ability to convert straight microwave antennas into omni-directional emitters for short-wavelength SWs will be a key to cellular nonlinear networks and integrated magnonics. Subsequently, it is now interesting to explore a variety of two-dimensional binary magnonic crystal and in this context the SW dynamics of embedded nanodots in a periodic ferromagnetic antidot lattice (annular antidot) structure has not been studied yet. The annular antidots represent a natural extension of the antidot geometry, in which each hole (antidot) contains a nanodot (central nanomagnet), which is separated from the antidot lattice by a nonmagnetic gap.

Here, we report the fabrication of two-dimensional bi-structure magnonic crystal in the form of embedded nanodots in a periodic $Ni_{80}Fe_{20}$ (Py) antidot lattice structure by focused ion-beam lithography. The static and dynamic responses in magnetic properties in Py annular antidot arrays are studied using time-resolved magneto-optic Kerr effect (TR-MOKE), magnetic force microscopy (MFM) imaging and micromagnetic simulations. The optically induced SW spectra show a remarkable variation with bias magnetic field. Multimode spectra are observed consisting of quantized SW mode, centre mode, edge mode, extended SW mode and backward volume (BV) mode due to collective interactions between the dots and antidots as well as the annular elements within the whole array. To gain a deeper understanding of the various modes in annular antidot samples, we study spin wave dynamics of pure dot lattices (DL) and pure antidot lattices (ADL) using micromagnetic simulations [39-40]. The pure DL shows three SW modes which are coherent precession of the centre and BV modes of the dots over the entire lattice. The pure ADL shows multiple modes which are mainly standing SW mode, BV mode, edge mode and quantized SW mode. Thus, the combined collective behaviour of both the DL and ADL is responsible for generating multiple frequency modes in the annular antidot lattice. Numerical simulations qualitatively reproduce the experimental results, and simulated mode profiles reveal the spatial distribution of the SW modes and internal magnetic fields responsible for these observations. MFM imaging shows the static magnetic configuration, which is well reproduced by micromagnetic simulations.

## 2. Sample fabrication

The periodic arrays of annular antidots arranged in square lattice symmetry were fabricated by a combination of e-beam evaporation (EBE) and focused ion beam (FIB) lithography as shown in Figure 1(a). Here, Py nanodots with 170 nm diameter are placed at the centre of holes with diameter 360 nm, which are periodically arranged on a square lattice with lattice constant $a$ = 480 nm. At first, 15 nm thick film of Py is deposited on top of silicon (Si) [100] substrate using e-beam evaporation in an ultrahigh vacuum chamber at a base pressure of $2\times10^{-8}$ torr. The film was then immediately transferred to a sputtering chamber for the deposition of a capping layer of 5 nm thick $SiO_2$ on top of the Py film to avoid degradation from natural oxidation and exposure to high power laser during optical pump-probe experiments in air. Deposition of $SiO_2$ was done by rf sputtering at a base pressure of $2\times10^{-7}$ torr, Ar pressure of 5 mtorr and using rf power of 60 W at a frequency of 13.56 MHz. In the next step, the annular antidot arrays are fabricated on the blanket Py film by using liquid $Ga^+$ ion beam lithography (Auriga-Zeiss FIB-SEM microscopes). The optimal values of voltage and current for milling are found to be 30 keV and 5 pA, respectively. In FIB, the spot size used for milling needs to be optimized to ensure that the material that has been retained after patterning (Py in this case) does not have excessive $Ga^+$ ion implantations. The spot size is a function of the beam current. We have used a beam current of 5 pA that gives sufficient etch rate yet limits the spot size to around 50 nm. The thickness of the Py film (15nm) is smaller than the stopping range of $Ga^+$ ions at 30 keV, which ensures that the ions stop within the Si layer underneath the Py film, which has been verified by atomic force microscopy (AFM) measurement.

## 3. Experimental details

The TR-MOKE microscope used in our investigation is based upon two-colour collinear optical pump-probe geometry [41]. Here, the second harmonic ($\lambda_b$ = 400 nm, 80 MHz, 10 mW, pulse width = 100 fs) of the fundamental beam of a mode locked Ti-Sapphire laser (Tsunami, Spectra physics) is used as pump to create hot electrons causing a modification of spin population. A part of the fundamental beam ($\lambda_a$ = 800 nm, 2mW, pulse width = 80 fs) is used to probe the time varying polar Kerr rotation from the sample. A delay stage situated at the probe path is used to create the necessary time delay between the pump-beam and the probe-beam with temporal resolution of about 100 fs limited by the cross-correlation of the pump and the probe beams. Finally, both the beams are combined together and focused at the centre of array by a microscope objective (N.A. = 0.65) in a collinear geometry. The probe-beam of spot size of about 800 nm diameter is tightly focused and overlapped with the slightly defocused pump-beam having larger diameter (~1 μm), at the centre of the array. Under this condition the probe can collect information from the uniformly excited part of the sample. A static magnetic field is applied at a small angle (~15°) to the sample plane, the in-plane component of which is defined as the bias field $H$. The magnitude of in-plane component of this field has to be large enough to saturate the magnetization. The time varying polar Kerr rotation is measured at room temperature by using an optical balance detector and a lock-in amplifier in a phase sensitive manner. The pump beam is modulated at 2 kHz frequency which eventually is being used as the reference frequency of the lock-in amplifier. This detection technique completely isolates the Kerr rotation and reflectivity signals. The measurement time window of about 2 ns used in this experiment is determined by the number of scan points and the integration time of the lock-in amplifier for each scan point. Nevertheless, this 2 ns time window is found to be sufficient to resolve the SW frequencies for this sample.

## 4. Results and discussions
### A. Sample characterization

A scanning electron microscope (SEM) is used to determine the actual sizes of the nanostructures. A representative SEM image of an annular antidot array is shown in Figure 1(a), which shows well-ordered array of dots at the centre of antidot structure and the bias field orientation is also drawn on the image. The dotted and solid lines show the circumferences of the antidot and the dot structure, respectively, which are used in micromagnetic simulations. The x and y axes of a Cartesian frame of reference along with the direction of the bias field is shown in the figure 1(a). The diameter (D) of the antidot is 360.0±3.5 nm, while the dot structure has diameter (d) of 170.0±1.5 nm. The edge-to-edge separation (s) of the adjacent annular antidot cells is measured to be 120.0±1.0 nm along the x and y directions. The EDX spectrum in Figure 1(b) shows the elemental composition of the sample having 82% of Ni and 18% of Fe in the sample, which is close to the nominal composition. A strong peak of Si is also present there signifying the substrate composition. We have omitted $O_2$ at the time of EDX, which is very less in composition. The AFM image in Figure 1(c) shows the topography of the annular antidot sample and the depth profile corresponding to the dotted line of AFM image can be studied from Figure 1(d). The completion of ion milling within the Si layer underneath the Py film (15 nm) is confirmed by depth profile, which shows depth ~25 nm. Figure 1(e) shows the MFM image of the sample, taken at remanent state shows clear magnetic contrast at the dot edges and magnetic regions between the antidot channel. The opposite MFM magnetic contrast shows the magnetic and nonmagnetic (milled) regions of the patterns. Some magnetic contrast, in the milled regions is due to the edge deformations near the boundaries of the dot and antidot edges. To gain a deeper understanding of the micromagnetic configuration giving rise to the magnetic contrast observed with MFM, we performed micromagnetic simulation as shown in Figure 1(f). The magnetic contrast (bright region) near the edges of the dot and antidot regions shows good agreement between simulation and experiment.

## B. Optical characterizations

Figure 2(a) shows a measured time-resolved Kerr rotation data for the annular antidot array with field $H$ = 1080 Oe having three different regimes within the nanosecond time window. After the pump pulse interacts with the sample, ultrafast demagnetization occurs within first 500 fs (Region-I). This is followed by two different relaxation processes, a fast relaxation (~20 ps, Region-II) and a slow relaxation (~100 ps). The spins deviate from the equilibrium state and start to precess around the effective magnetic field causing time resolved Kerr trace (Region-III) to appear as an oscillatory signal on the slowly decaying part of the trace. Figure 2(b) shows the background subtracted precessional Kerr rotation data obtained after removing the negative delay and ultrafast demagnetization. It contains clear signature of damped sinusoidal nature due to the dephasing of multimodes present in the magnetization dynamics. The time-resolved data corresponding to the reflectivity of the array is shown in Figure 2(c) to show that the precessional oscillation is not related to a possible breakthrough of reflectivity signal in the Kerr rotation signal. Fast Fourier transform (FFT) is performed over the background subtracted Kerr rotation data using rectangular window to obtain the frequency spectra of the SWs (shown in Figure 2(d)). Rich SW spectra, which varied systematically with the bias field variation ($H$ = 1080 Oe, 840 Oe, 720 Oe, 620 Oe and 520 Oe) are obtained for the array. In Figure 2(d), broad band of SW modes in the range 3.5 GHz to 14.7 GHz with different peak intensities are shown for the highest bias field ($H$ = 1080 Oe). A number of peaks appear in the spectra, out of which seven modes are identified as magnetic modes whereas some other peaks appear in the higher and lower frequency sides with relatively lower intensities, mainly due to the presence of small amount of external noise in the signal. Bias magnetic field dependence of the SW spectra in the experiment is used to identify the magnetic modes, whereas the peaks appearing due to non-magnetic background do not vary with the magnetic field, and are not included in the analysis. Higher frequency SW modes are relatively intense (mainly modes 2, 3 and 4) with maximum increase in broadening of about 0.6 GHz as compared to the simulation result. The whole spectra gradually shift to the lower frequency range with decreasing bias field. Interestingly we have found that another SW mode (marked with * in Figure 2(d)) rises at the tail of mode 3 of the bias fields 520 Oe and 620 Oe, which merges with mode 3 at H = 720 Oe. For H = 620 Oe, modes 6 and 7 appear to merge together resulting a line broadening of almost 1 GHz.

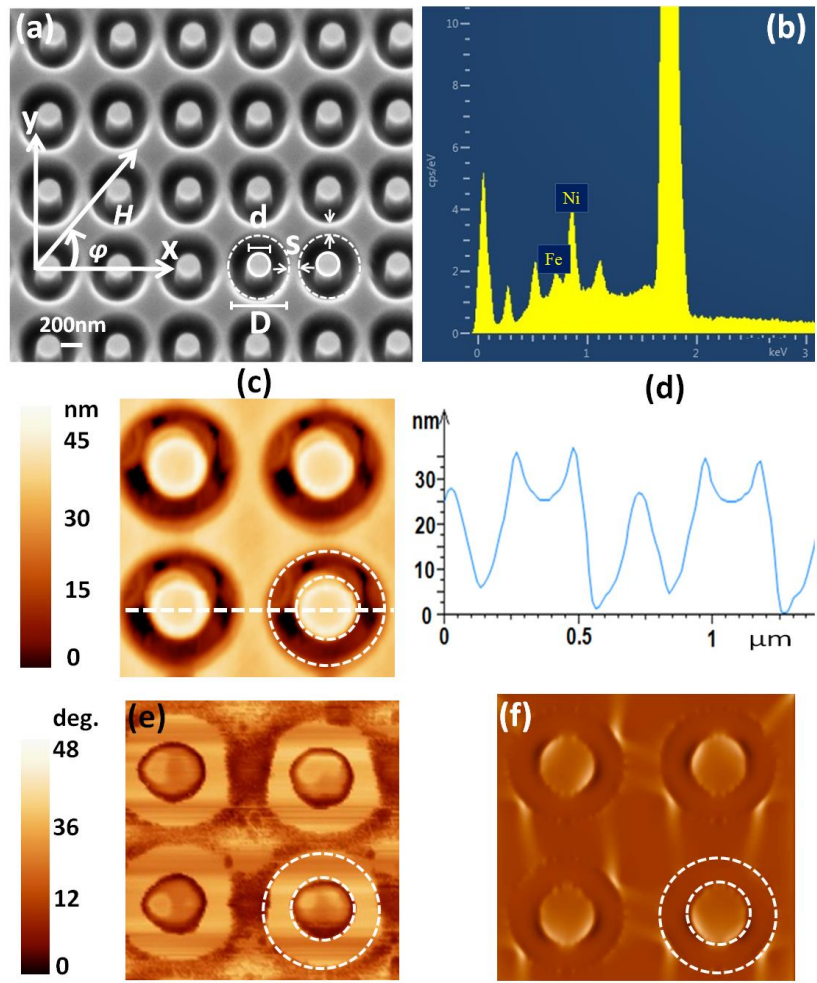

FIGURE 1(a) Scanning electron micrograph of 15 nm thick Py annular antidot lattice with inner dot diameter (d) = 170±1.5 nm, annular antidot diameter (D) = 360±3.5 nm and edge-to-edge separation of annular antidot (s) = 120±1 nm and the geometry of the applied magnetic field of the measurement. Here $\varphi$ is the angle between applied bias field and x coordinate which was kept at 0° during the measurement. (b) EDX spectrum of Py annular antidot sample. (c) AFM image, (d) corresponding line scan profile along the dotted line of annular antidot array and (e) experimental and (f) simulated MFM images of the annular antidot array, respectively. The color bars for AFM and MFM images are shown at the left side of the images. The white dotted circles show that features of dot and antidots in the AFM and MFM images are comparable.

### C. Micromagnetic Simulations

We have performed numerical calculations using Object Oriented Micro Magnetic Framework (OOMMF) [39] on our patterned samples to account the experimental results presented earlier. In the simulation, the equilibrium magnetic configuration is first prepared at the appropriate bias magnetic field followed by application of a perpendicular pulsed magnetic field for excitation of magnetization precession in the system. The experimental condition of optically triggering the magnetization dynamics, is thus reliably reproduced in simulations. The details of the simulation can be found elsewhere [42]. The simulations are performed on a sample volume of $1900 \times 1900 \times 15$ nm$^3$ consisting $4 \times 4$ elements after applying two-dimensional periodic boundary conditions (2-D PBC) [43]. The samples are discretized into cuboidal cells of volume $5 \times 5 \times 15$ nm$^3$ where the lateral cell size is kept below the exchange length of Py (5.2 nm). We also performed some test simulations with larger arrays containing $7 \times 7$ annular antidots to check if the artificial boundaries of the simulated lattices for these combinations of dot and antidot nano-structures can affect the frequencies of SW modes. We observe that the boundaries do not affect the mode frequencies significantly, but they do affect the relative intensities of the modes in the FFT spectra with varying bias field values which verify the experimental observations. The shapes of the annular antidots used in simulation are derived from the SEM image. The material parameters used in the simulations are gyromagnetic ratio, $\gamma = 17.5$ MHz Oe$^{-1}$, magneto-crystalline anisotropy, $H_K=0$, saturation magnetization, $M_s = 860$ emu cc$^{-1}$ and exchange stiffness constant, $A = 1.3 \times 10^{-6}$ erg cm$^{-1}$. The $M_s$ value is extracted by experimentally measuring the precessional frequency ($f$) as a function of bias magnetic field ($H$) (Figure 3(a)) of a continuous Py thin film of 15 nm thickness deposited under the same condition as the arrays and by fitting the data with Kittel formula (Equation 1) for the uniform precession mode:

$$f = \frac{\gamma}{2\pi}\sqrt{(H+H_K)(H+H_K+4\pi M_s)} \quad (1)$$

Using the above material parameters in the simulation, the experimental mode frequencies are qualitatively reproduced. The average deviation of simulated data from experimental value in terms of mode frequencies is about ± 6%. However, the relative mode intensities could not be correctly reproduced due to various factors including the lack of inclusion of precise edge roughness, and statistical differences in antidot size of the real sample as well as the simulation temperature of T = 0 K as opposed to the experimental temperature of T = 295 K. The effect of milling of the sample causes edge roughness, resulting in the modification of edge modes and introduction of additional localized modes through randomization of magnetostatic stray fields.

Figure 2(e) shows the simulated SW spectra for the annular ADL, with different bias field values. For $H$ = 1080 Oe, seven modes are present. These modes shift to the lower frequency regime with the lowering of bias field values. The relative peak intensities between mode 3 and mode 4 changes drastically with the bias field variation. A notch appears in mode 3 at 1080 Oe, which gains in amplitude at lower bias fields as observed in experiments too for $H$= 620 and 520 Oe. The signal to noise ratio decreases for the lowest field value due to the reduction of field strength. It results the broadening of peaks as well as appearance of some low intensity peaks in the higher frequency side which are not at all present in the simulated SW spectra. In Figure 3(b) SW frequencies are plotted with respect to bias field that are extracted from experimental and simulated FFT spectra. The frequencies corresponding to modes 4 and 5 are well fitted with Kittel formula, while the higher frequency modes (i.e. modes 1, 2, 3) as well as lower frequency modes do not follow the Kittel formula. The $M_s$-values obtained from the Kittel fit of modes 4 and 5 are 830 emu cc$^{-1}$ and 490 emucc$^{-1}$, respectively, while the other magnetic parameters are found to be similar to the 15 nm Py thin film values. To have deeper understanding of bias field dependence of SW modes for the annular ADL, we have further simulated the magnetization dynamics of pure antidot and pure dot arrays. We have considered a pure antidot lattice (ADL) having 4×4 elements along with 2-D PBC arranged in square symmetry with hole diameter of 360 nm, edge-to-edge separation of 120 nm and thickness of 15 nm. Figure 3(c) shows the bias field dependence of the simulated precessional frequency of SW modes from the pure ADL. Total 7 modes are present in the higher field regime. Mode 5 is fitted with the Kittel formula and $M_s$ value obtained from the fitting is 480 emucc$^{-1}$, which is close to the value obtained for mode 4 of the annular ADL. Mode 6 splits at the lower field regime (620 Oe and 520 Oe) as shown in Figure 3(c). This is probably because of the large influence on the lower frequency modes because of the competition between bias field and magnetic stray field generated from the unsaturated spins of the antidot edges. Beside these we have also simulated the SW mode profile for an array of pure dot lattice having 4×4 elements arranged in square symmetry with diameter 170 nm and edge to edge separation 310 nm. The thickness has been considered to be 15 nm same as the annular ADL. The Kittel fit to the bias field dependence of SW mode frequencies yields $M_s$ value of 670 emucc$^{-1}$ for mode 1. Comparing the $M_s$ values of annular ADL with those of the pure ADL and DL, we find that mode 4 of annular ADL has a similar value to the mode 5 of pure ADL, while mode 3 of the annular ADL has $M_s$ value, significantly larger than mode 1 of pure DL. Hence, it is non-trivial to understand the nature of the modes of the annular ADL from the bias field dependence of the mode frequencies and more detailed analyses of the mode profiles are carried out as below.

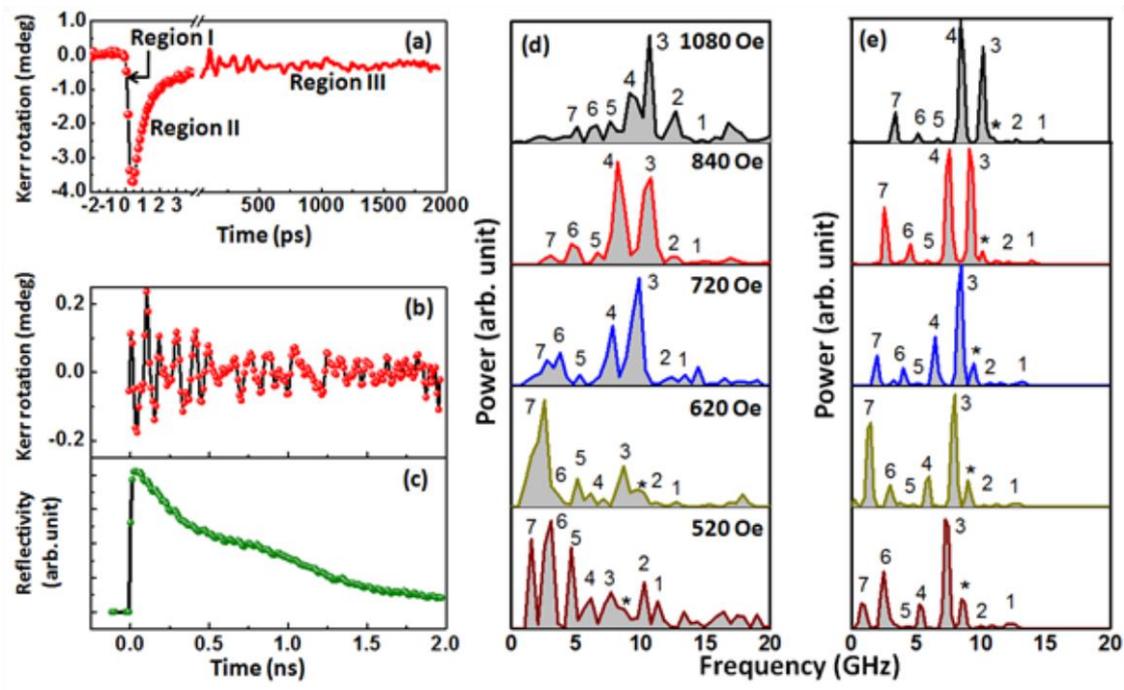

FIGURE 2: Time-resolved (a) Kerr rotation (full scale data) (b) background subtracted Kerr rotation and (c) reflectivity data are shown for the sample at $H$ = 1080 Oe at $\varphi$ = 0°. Power vs. frequency spectra of the sample with varying bias fields obtained from using (d) time-resolved MOKE and (e) micromagnetic simulations.

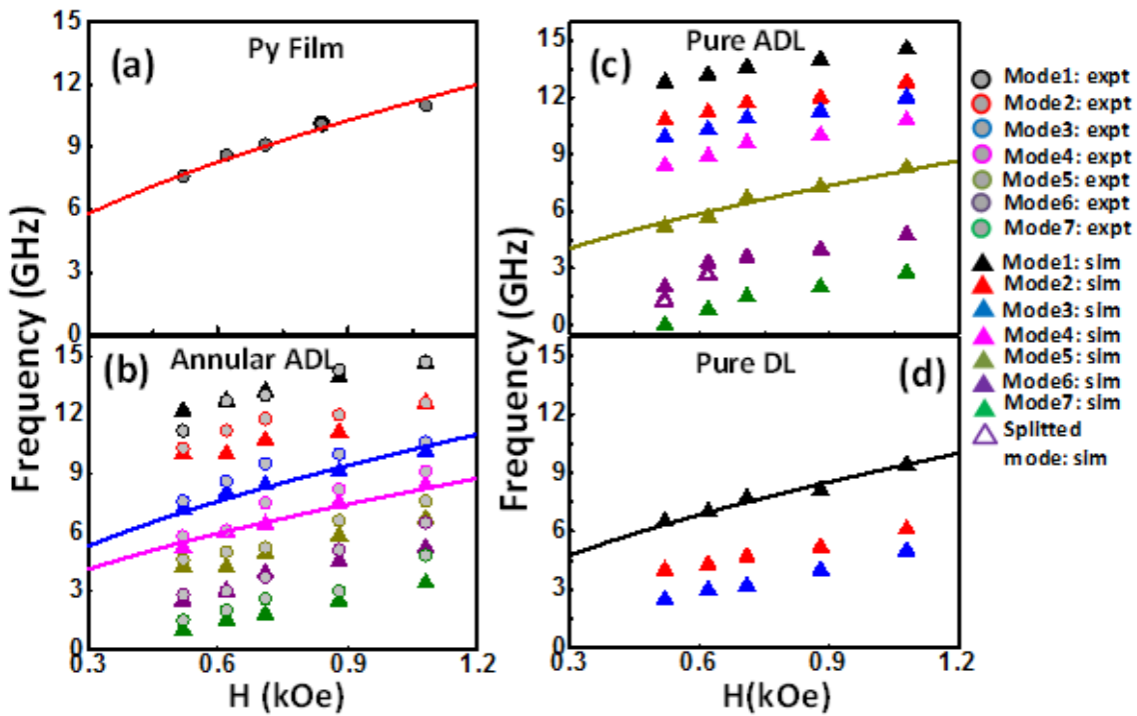

FIGURE 3: Precessional frequencies of different SW modes for (a) Py thin film, (b) annular antidot sample, (c) pure antidot lattice and (d) pure dot lattice (circular symbols: experimental data, triangular symbols: micromagnetic simulation results, solid line: Kittel fit) are plotted as a function of bias field $H$.

### D. Power and phase profiles

We further simulate the power and phase maps using a home-grown Matlab code [44] for various collective modes for annular ADL, pure ADL and pure DL, observed in micromagnetic simulation where the bias fields of 1080 Oe and

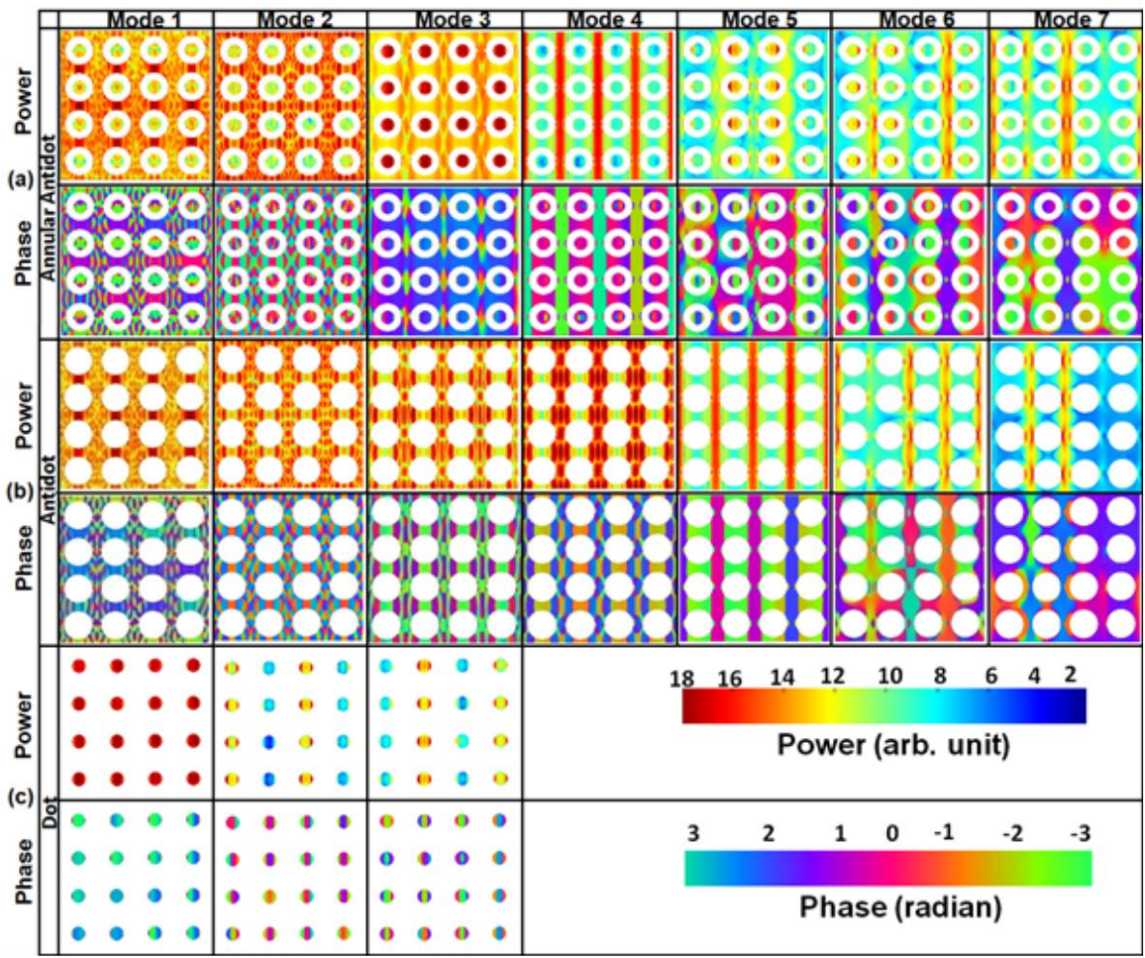

FIGURE 4: The power and phase maps for different resonant modes (as shown in Fig. 2 (e)), of (a) annular antidot lattice, (b) antidot lattice and (c) dot lattice at a bias field of 1080 Oe. The colormaps for the power and phase distributions are shown at the bottom right corner of the images.

520 Oe are applied along the horizontal edges of the arrays as shown in Figures 4 and 5, respectively. The spatial profiles of the power and phase information for various resonant modes are obtained by fixing one of the spatial coordinates in the space and time-dependent magnetization and then by performing a discrete Fourier transform with respect to time domain. For annular ADL, the highest frequency mode (14.7 GHz) corresponds to a mixed backward volume (BV, quantization number = $n$) – Damon Eshbach (DE, quantization number = $m$) mode of the internal dot with quantization number $n = 3$, $m = 3$ as shown in Figure 4(a). Within the antidot, the mode is quantized in nature with $n' = 9$ where the power is mainly concentrated in the intermediate channel between two consecutive holes. Mode 2 (12.7 GHz) of annular ADL is a BV-like standing SW mode with quantization numbers $n = 6$ within the dot and $n' = 7$ within the antidot. Mode 3 (10.2 GHz) corresponds to the centre mode of the dots distributed uniformly through the whole array. The nature of the SWs within the antidot is quantized BV-like mode with $n' = 3$. Mode 4 (8.5 GHz) shows that the maximum power is concentrated along the vertical channel between the antidots. This is called the extended DE-like SW mode of the annular ADL where the individual dots contain very small power only at their edges. Mode 5 (6.7 GHz) corresponds to the edge mode of the antidot while the dots show BV-like mode with $n = 3$ but with asymmetric power and phase distribution due to the interaction between the antidot and dot edges. In mode 6 (5.2GHz) the edge modes of the antidots in the alternating channel start to interact, while the dots show BV-like mode with $n = 3$ but with reversed asymmetry as in the case of mode 5. In mode 7 (3.5 GHz) the edge modes interact very strongly through the central part of the antidot creating almost an extended mode through that channel, while the dots do not show any significant precessional amplitude.

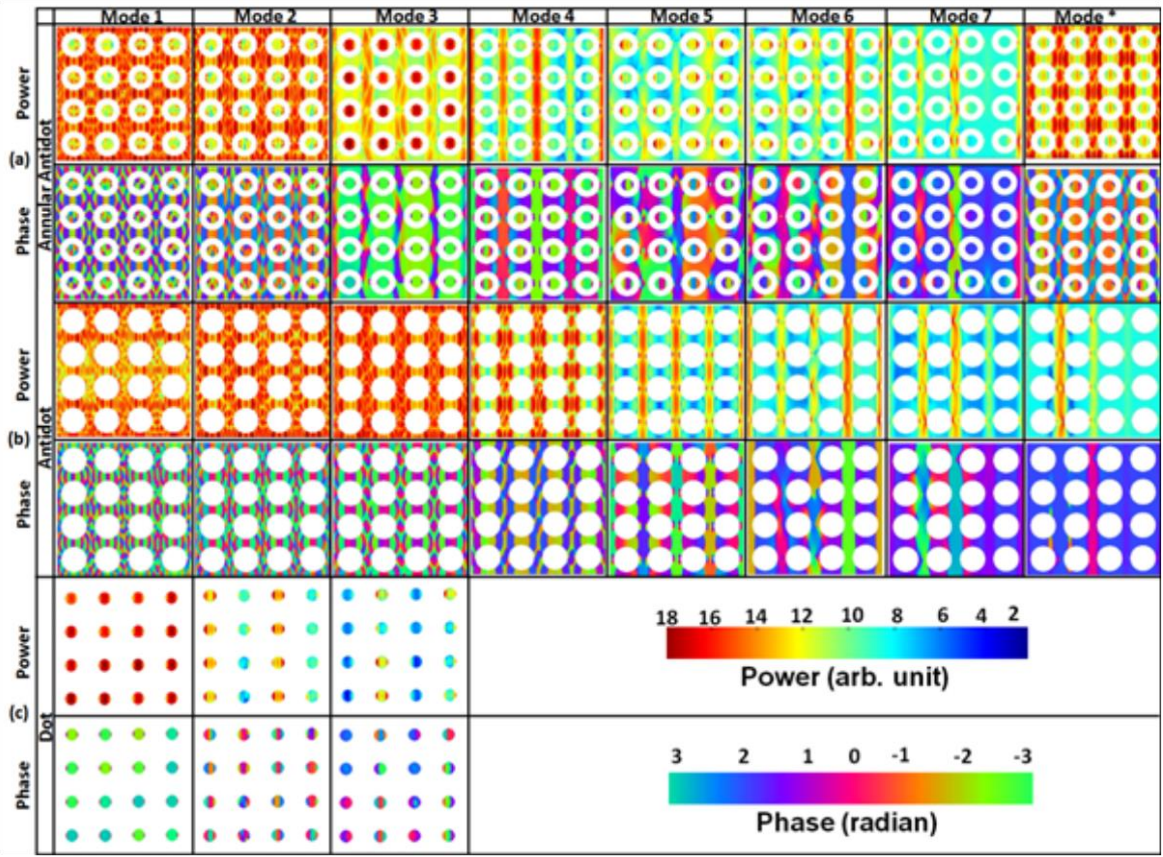

FIGURE 5: The power and phase maps for different resonant modes (as shown in Fig. 2 (e)) of (a) annular antidot lattice, (b) antidot lattice and (c) dot lattice at a bias field of 520Oe. The colormaps for the power and phase distributions are shown at the bottom right corner of the images.

TABLE I. Comparison of Spin-wave mode-frequencies of annular antidot lattice (annular-ADL), pure antidot lattice (ADL) and pure dot lattice (DL)

| Bias field (Oe) | Mode no. | Frequency of Modes in GHz | | | | |
|---|---|---|---|---|---|---|
| | | Thin film | Annular antidot lattice (Annular ADL) | | Antidot lattice (ADL) | Dot lattice (DL) |
| | | | Exp. | Sim. | | |
| 1080 | 1. | 11.0 | 14.7 | 14.7 (Mixed BV & DE) | 14.6 (BV) | 9.4 (Centre) |
| | 2. | - | 12.6 | 12.7 (BV(ADL)&BV(DL)) | 12.8 (BV) | 6.1 (BV) |
| | 3. | - | 10.6 | 10.2 (BV(ADL) & Centre(DL) | 12.0 (BV) | 5.0 (edge mode) |
| | 4. | - | 9.1 | 8.5 (DE(ADL) & Edge(DL) | 10.8 (BV) | - |
| | 5. | - | 7.6 | 6.7 (Edge(ADL) & BV(DL) | 8.3 (Extended DE) | - |
| | 6. | - | 6.5 | 5.2 (Edge(ADL) & BV(DL) | 4.8 (Edge) | - |
| | 7. | - | 4.8 | 3.5(Edge(ADL)) | 2.8 (Edge) | - |
| 840 | 1. | 10.1 | 14.3 | 13.9 (Mixed BV & DE) | 14.0(BV) | 8.1 (Centre) |
| | 2. | - | 12.0 | 11.1 (Mixed BV & DE) | 12.0(BV) | 5.2(BV) |
| | 3. | - | 10.0 | 9.1 (BV(ADL) & Centre(DL) | 11.3(BV) | 4.0(BV) |
| | 4. | - | 8.2 | 7.5 (DE(ADL) & Edge(DL) | 10.0(BV) | - |
| | 5. | - | 6.6 | 5.8 (Edge(ADL) & BV(DL) | 7.3(DE) | - |
| | 6. | - | 5.1 | 4.5 (Edge(ADL) & BV(DL) | 4.0(Edge) | - |
| | 7. | - | 3.0 | 2.5 (Edge(ADL)) | 2.0(Edge) | - |
| 710 | 1. | 9.1 | 13.0 | 13.2 (Mixed BV & DE) | 13.6(BV) | 7.7 (Centre) |
| | 2. | - | 11.8 | 10.7 (Mixed BV & DE) | 11.7(BV) | 4.7(BV) |
| | 3. | - | 9.5 | 8.4 (BV(ADL) & Centre(DL) | 10.9(BV) | 3.2(BV) |
| | 4. | - | 7.5 | 6.4 (DE(ADL) & Edge(DL) | 9.6(BV) | - |
| | 5. | - | 5.2 | 4.9 (Edge(ADL) & BV(DL) | 6.7(DE) | - |
| | 6. | - | 3.7 | 3.9 (Edge(ADL) & BV(DL) | 3.6(Edge) | - |
| | 7. | - | 2.6 | 1.8 (Edge(ADL)) | 1.5(Edge) | - |
| 620 | 1. | 8.6 | 12.7 | 12.7 (Mixed BV & DE) | 13.2(BV) | 7.0 (Centre) |
| | 2. | - | 11.2 | 10.0 (Mixed BV & DE) | 11.2(BV) | 4.3(BV) |
| | 3. | - | 8.6 | 8.0 (BV(ADL) & Centre(DL) | 10.3(BV) | 3.0(BV) |
| | 4. | - | 6.1 | 6.0 (DE(ADL) & Edge(DL) | 8.9(BV) | - |
| | 5. | - | 5.0 | 4.2 (Edge(ADL) & BV(DL) | 5.7(DE) | - |
| | 6. | - | 3.0 | 3.0 (Edge(ADL) & BV(DL) | 3.3& 2.7 (Edge) | - |
| | 7. | - | 2.0 | 1.5 (Edge(ADL)) | 1.3 &0.8 (Edge) | - |
| | * | - | 10.2 | 9.0 (BV(ADL) & BV(DL)) | - | - |
| 520 | 1. | 7.6 | 11.2 | 12.2 (Mixed BV & DE) | 12.8(BV) | 6.5 (Centre) |
| | 2. | - | 10.3 | 10.0 (Mixed BV & DE) | 10.8(BV) | 4.0(BV) |
| | 3. | - | 7.6 | 7.2 (BV(ADL) & Centre(DL) | 9.9(BV) | 2.5(BV) |
| | 4. | - | 5.8 | 5.2 (DE(ADL) & Edge(DL) | 8.4(BV) | - |
| | 5. | - | 4.6 | 4.3 (Edge(ADL) & BV(DL) | 5.2(DE) | - |
| | 6. | - | 2.8 | 2.5 (Edge(ADL) & BV(DL) | 2.0 (Edge) | - |
| | 7. | - | 1.5 | 1.0 (Edge(ADL)) | 1.3; 0.3 (Edge) | - |
| | * | - | 8.7 | 8.5 (BV(ADL) & BV(DL)) | - | - |

To have deeper understanding of the collective interactions between the dots and antidots on the overall dynamics of the annular ADL, we have further simulated the power and phase profiles of arrays of constituent dot lattice (DL) and antidot lattice (ADL), separately. Figure 4(b) shows that the highest frequency mode (14.6 GHz) of the ADL is BV-like quantized mode with $n' = 9$. This mode sustains its behaviour in the annular ADL with almost the same frequency but different intensity. Mode 2 (12.8 GHz) of the ADL is also similar to that of annular ADL. Modes 3 (12.0 GHz) and mode 4 (10.8 GHz) of the ADL are also BV-like quantized modes having quantization numbers $n' = 5$ and 3 respectively, but these two modes no longer exist in the annular ADL and instead a new collective mode (10.2 GHz) appears due to the strong magnetostatic interactions between the unsaturated spins at the edges of dots and antidots. Mode 5 (8.3 GHz), on the other hand, is the extended mode in the DE geometry, which appears as mode 4 (8.5 GHz) in the annular ADL with slightly varying frequency and power as also confirmed from the Kittel fit to the frequency vs bias magnetic field result. Mode 6 (4.8 GHz) and mode 7 (2.8 GHz) are the interacting edge modes of the ADL which appears with different power distributions in mode 5 (6.7 GHz), mode 6 (5.2 GHz) and mode 7 (3.5 GHz) of the annular ADL. The three resonant modes appeared in the FFT spectra of the DL are mainly the centre mode (mode 1), BV-like standing wave mode with $n = 3$ (mode 2) and the edge mode (mode 3) of the dots in the lattice. Mode 1 (9.5 GHz) of the DL appears in annular ADL as mode 3 (10.2 GHz) with almost same power distribution, while modes 2 (6.1 GHz) and 3 (5.0 GHZ) of the DL appear as modes 5

(6.7 GHz) and 6 (5.2 GHz), respectively of the annular ADL. However, a number of modes of the annular ADL neither appear in the pure DL nor in pure ADL and are understood to be new collective modes originated from the magnetic interactions between the constituent DL and ADL making the annular ADL.

For $H = 520$ Oe (Figure 5), the modes of annular ADL are almost identical to those observed for $H = 1080$ Oe except for one mode (Figure 5(a)) marked by the asterisk (*). This additional mode is a combination of BV-like standing wave mode in both the dot ($n = 3$) and the antidot ($n' = 3$) similar to mode 4 (Figure 5(b)) of the ADL and mode 2 of the DL (Figure 5(c)). Some additional low frequency modes also appear in the pure DL and ADL, which are not observed in the annular ADL. All other modes of annular antidot are modified due to the interaction as compared to pure dot

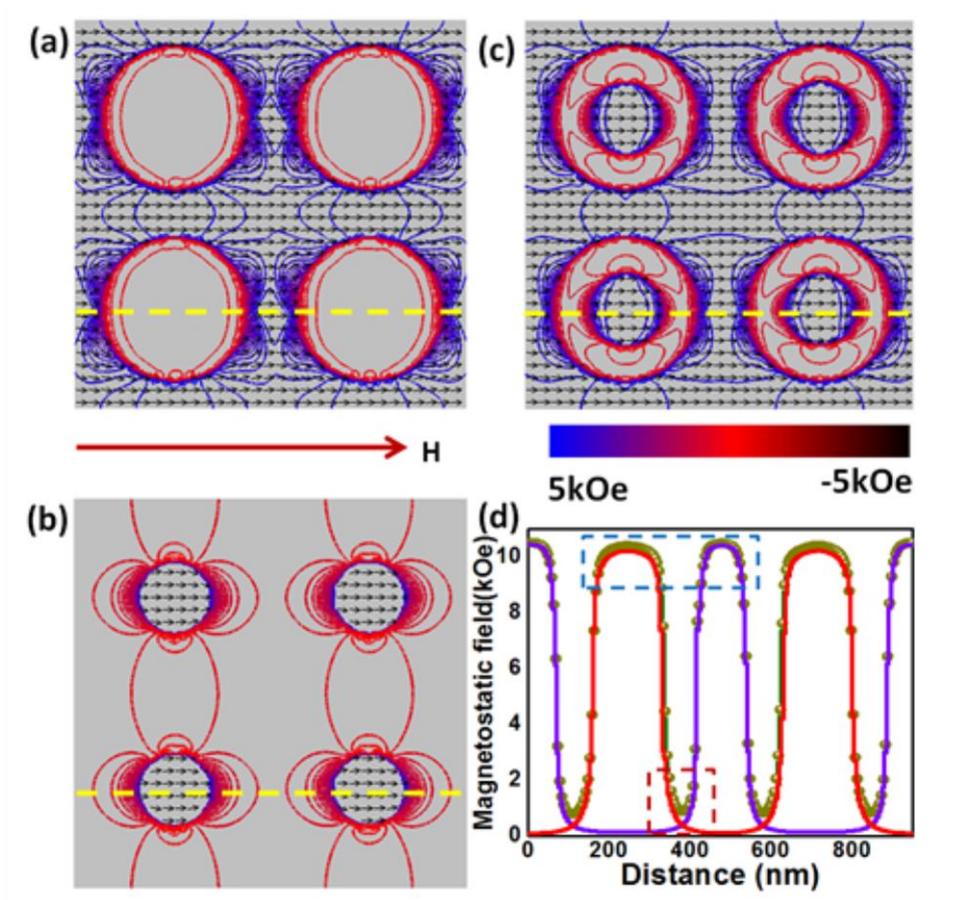

FIGURE 6: Contour maps of simulated magnetostatic field distributions (x-component) are shown for (a) antidot lattice (ADL), (b) dot lattice(DL) and (c) annular antidot lattice(annular ADL) samples at $H = 1080$ Oe at $\varphi = 0°$. The arrows represent the magnetization states of the structures, while the strength of the stray fields is represented by the color bar at the bottom of the sample. Comparison of the simulated magnetostatic field distributions in ADL, DL and annular ADL is shown in (d) taken along the dotted lines from samples.

and pure antidot structure.

Different SW modes are listed in Table I for the annular antidots in comparison with pure antidot and pure dot lattices for better clarity in the relationships between them and variation with bias field.

### E. Magnetostatic field distribution

To understand the origin of the SW modes and the interesting features, we calculate the magnetostatic field distribution of the systems. Figure 6 (a), (b) and (c) show the ground state of pure ADL, pure DL and annular ADL respectively obtained using LLG micromagnetic simulation for $H = 1080$ Oe. One can readily observe that in case of pure ADL (Figure 6(a)) the field lines are very dense between the two edges of the neighbouring antidots, while they extend by a very small amount inside the holes.

In the case of pure DL (Figure 6(b)) the stray field extends from the dot edges but they do not interact due to the large enough separation between the edges of the neighbouring dots. On the contrary, in the case of annular ADL (Figure 6(c)) the field lines become less dense between the edges of the neighbouring antidots and they extend significantly inside the holes to interact with the dots sitting at the centre of the holes. This is also apparent from the linescans of the magnetostatic field as shown in Figure 6(d). In presence of bias field, the internal field values for pure ADL and DL are 10.4 and 10.2 kOe, respectively. In case of annular ADL the internal field strength increases by about 200 Oe inside each dot. This clear improvement of internal field strength is achieved due to the strong interaction between the two edges of the dot and antidot. However, the internal field within the channel between two antidots remains almost same (10.4 kOe). The stray magnetic field at the centre between two dots (antidots) in case of pure DL (ADL) are negligible, while that between the dot and antidot in case of the annular ADL is substantially large (about 0.8kOe). This stronger interacting field and elevated internal fields lead towards the additional collective modes in the annular ADL and a change of their frequencies even when the nature of the modes is qualitatively retained.

### 5. Conclusion

We have fabricated two-dimensional Py annular antidot lattices, where nanodots with diameter 170 nm are placed at the centre of holes of diameter 360 nm and edge-to-edge separation of annular antidot regions is 120 nm, arranged in a square lattice with lattice constant of 480 nm by using electron beam evaporation and focused ion beam lithography. We have investigated the time-resolved magnetization dynamics in this sample by varying the external bias magnetic fields using TR-MOKE microscopy. The optically induced SW spectra show multiple resonant modes due to the collective SW dynamics of the interacting dot and antidot lattices. Interestingly, for fixed annular antidot structure, the magnetization dynamics is markedly sensitive to the variable external bias field values due to complex spatial distributions of the internal and stray magnetic field. The dynamics has also been simulated by a time-dependent micromagnetic simulation method to obtain the power spectra. The power and phase profiles of the resonant modes have been numerically calculated to get an extensive picture of the SW dynamics profile. The resonant modes of magnetization show significant variation with the external bias magnetic fields. We observe multiple frequency modes in the annular antidot sample. To get a good understanding of these modes we have also performed micromagnetic simulations on pure DL and ADL separately. The central nanomagnet (dot) shows three modes which are coherent precession of the edge mode, BV-like standing SW mode and centre modes of the dots over the entire lattice. The antidot region shows multiple modes which are mainly standing SW modes of purely BV or mixed BV-DE origin, extended mode and edge mode. Thus, the collective behaviour of both the DL and ADL is modifying the spin wave dynamics of the whole annular ADL sample and are responsible for generating multiple frequency modes. Simulated internal field and stray field profiles throw further insight to the observed SW modes. Observation of new collective SW modes, which are neither observed in the pure DL nor in pure ADL, primarily due to the strong interaction between the dots and the antidots is significant for building future magnonic and spintronic devices based on such patterned structures.

### Acknowledgement

The authors acknowledge the SGDRI (UPM) project of IIT Kharagpur for support. AB acknowledges S. N. Bose National Centre for Basic Sciences for financial support (Project No. SNB/AB/12-13/96). SM and AD acknowledge DST for INSPIRE fellowship, while SC acknowledges S. N. Bose National Centre for Basic Sciences for senior research fellowship.